\documentclass{aa}
\usepackage{graphicx}
\usepackage{txfonts}

\begin{document}
\title{Metallicity as a criterion to select H$_2$ bearing Damped Lyman-$\alpha$ 
systems\thanks{Based on observations 
carried out at the European Southern Observatory (ESO) under prog. ID No. 65.O-0158, 65.O-0411,
67.A-0022, 67.A-0146, 69.A-0204, 70.B-0258, 072.A-0346  with 
UVES installed at the Very Large Telescope (VLT) on
Cerro Paranal, Chile}
}

\titlerunning{H$_2$ in DLA systems}

\author{P. Petitjean\inst{1,2} \and C. Ledoux\inst{3} \and P. Noterdaeme\inst{3}
\and R. Srianand\inst{4}}

\institute{Institut d'Astrophysique de Paris, CNRS - Universit\'e Pierre et Marie Curie, 
98bis Boulevard Arago, 75014, Paris, France $-$ \email{petitjean@iap.fr}
\and
LERMA, Observatoire de Paris, 61 Avenue de l'Observatoire, 75014, Paris, France 
\and
European Southern Observatory, Alonso de C\'ordova 3107, Casilla 19001, Vitacura, 
Santiago, Chile
\and
IUCAA, Post Bag 4, Ganesh Khind, Pune 411 007, India\\
}

\date{Received ; accepted}
 
  \abstract
   {}
   {We characterize the importance of metallicity on the presence of molecular hydrogen in 
damped Lyman-$\alpha$ (DLA) systems.}
   {We construct a representative sample of 18 DLA/sub-DLA systems with log~$N$(H~{\sc i})~$>$~19.5 at 
high redshift ($z_{\rm abs}$~$>$~1.8) with metallicities relative
to solar [X/H]~$>$~$-$1.3 (with [X/H] = log~$N$(X)/$N$(H)$-$log(X/H)$_{\odot}$ and X either Zn, S or Si). 
We gather data covering the expected wavelength range of redshifted H$_2$ absorption lines on all 
systems in the sample from either the literature (10 DLAs), the UVES-archive or new VLT-UVES 
observations for four of them. The sample is large enough to discuss for the 
first time the importance of metallicity as a criterion for the presence of molecular hydrogen in the 
neutral phase at high-$z$.
}
   {From the new observations, we report two new detections of molecular hydrogen in the systems at 
$z_{\rm abs} = 2.431$ toward Q\,2343$+$125 and $z_{\rm abs}= 2.426$ toward Q\,2348$-$011. 
We compare the H$_2$ detection fraction in the high-metallicity sample with the 
detection fraction in the overall sample from Ledoux et al. (2003). We show that the 
fraction of DLA systems with log~$f$~=~log~2$N$(H$_2$)/(2$N$(H$_2$)~+~$N$(H~{\sc i}))~$>$~$-$4
is as large as 50\% for [X/H]~$>$~$-$0.7 
when it is only $\sim$5\% for [X/H]~$<$~$-$1.3 and $\sim$15\% in the overall sample 
(with $-$2.5~$<$~[X/H]~$<$~$-$0.3). This demonstrates that the presence of molecular hydrogen 
at high redshift is strongly correlated with metallicity.
%
}
%
   {}


   \maketitle

\section{Introduction} 

Early searches for molecular
hydrogen in DLAs, though not systematic, have lead to either small values of or
upper limits on the molecular fraction of the gas.
For a long time, only the DLA at $z_{\rm abs}=2.811$ toward Q\,0528$-$250 was
known to contain H$_2$ molecules (Levshakov \& Varshalovich 1985, 
Foltz et al. 1988). Ge \& Bechtold (1999)
searched for H$_2$ in a sample of eight DLAs using the
MMT moderate-resolution spectrograph (${\rm FWHM}=1$ \AA ). Apart from the
detection of molecular hydrogen at $z_{\rm abs}=1.973$ and 2.338
toward, respectively, Q\,0013$-$004 and Q\,1232$+$082, they measured in
the other systems upper limits on $f$ in the range $10^{-6}-10^{-4}$.

A major step forward in understanding the nature of DLAs through their
molecular hydrogen content has recently been made possible by the
unique high-resolution and blue-sensitivity capabilities of UVES at the VLT. 
In the course of the
first large and systematic survey for H$_2$ at high redshift, we have searched
for H$_2$ in DLAs down to a detection limit of typically
$N($H$_2)=2\times 10^{14}$ cm$^{-2}$ (Petitjean et al. 2000,
Ledoux et al. 2003). Out of the 33 surveyed systems, eight had firm detections of 
associated H$_2$ absorption lines. Considering that three detections were already known from
past searches, H$_2$ was detected in $\sim 15$\% of
the surveyed systems. The existence of a correlation between metallicity
and depletion factor, measured as [X/Fe] (with X~=~Zn, S or Si) was demonstrated 
(see also Ledoux et al. 2002a) and 
the DLA and sub-DLA systems where H$_2$ was detected were usually among 
those having the highest metallicities.

However, the high metallicity end of the Ledoux's sample was biased
by the presence of already known detections.
Therefore, to investigate further this possible dependence with metallicity,
and derive what is the actual molecular content of the high-redshift
gas with highest metallicity, we have searched a representative 
sample of high metallicity DLAs for H$_2$. 
The number of H$_2$ measurements in systems with [X/H]~$>$~$-$1.3 is
twice larger in our sample compared to previous surveys.
We describe the sample and the observations in Section~2,
present two new detections of H$_2$ in Section~3 and the results of the survey
and our conclusions in Section~4. 

\section{Sample and Observations}

We have selected from the literature all $z>1.8$ DLAs and sub-DLAs systems ($\log N($H\,{\sc i}$)\ge 19.5$) with 
previously measured elemental abundances (e.g. Prochaska \& Wolfe 2001, Kulkarni \& Fall 2002) 
larger than [X/H]~$>$~$-$1.3 and accessible to UVES. The inclusion of sub-DLAs is 
justified by the fact that for $\log N($H\,{\sc i}$) > 19.5$ most of the hydrogen
is neutral (Viegas 1995). In addition, Ledoux et al. (2003) 
have shown that for log~$N$(H~{\sc i})~$>$~19.5, there is no correlation between the 
presence of H$_2$ and the H~{\sc i} column density. 
The fact that we include ALL known systems with these criteria
guarantees that the sample is representative of the population of DLA-subDLA.

We ended up with a sample of 15 DLAs and 3 sub-DLAs (with log~$N$(H~{\sc i})~=~19.7,
20.10 and 20.25) presented in Table~1. The H$_2$
content of ten of these systems had already been published (Srianand \& Petitjean 2001, 
Petitjean et al. 2002, Ledoux et al. 2002b, 2003, 2006b, Srianand et al. 2005 
and Heinm\"uller et al. 2006). Five of the eight remaining systems had 
data in the UVES archive (Q\,1209$+$093: Prog. 67.A-0146 P.I. Vladilo; Q\,2116$-$358: 
Prog. 65.O-0158 P.I. Pettini; Q\,2230$+$025: Prog. 70.B-0258, P.I. Dessauges-Zavadsky;
Q\,2243$-$605: Prog. 65.O-0411 P.I. Lopez; Q\,2343$+$125: Prog. 69.A-0204 and 67.A-0022,
P.I. D'Odorico). 
New observations of Q\,0216$+$080, Q\,2206$-$199, Q\,2343$+$125 and Q\,2348$-$011
have been performed with the Ultraviolet and Visible Echelle Spectrograph 
(UVES, Dekker et al. 2000) 
mounted on the ESO Kueyen VLT-UT2 8.2 m telescope on Cerro Paranal in Chile.
These new observations resulted in two new detections described in the next Section.
\\
The data for each of the eight QSOs 
were reduced using the UVES
pipeline 
which is available as a context of the ESO
MIDAS data reduction system (see e.g. Ledoux et al. 2003 for details). 
Standard Voigt-profile fitting methods were used for the analysis of metal lines
and molecular lines, when detected,
to determine column densities using the oscillator strengths compiled in 
Ledoux et al. (2003) for metal species and the oscillator strengths given 
by Morton \& Dinerstein (1976) for H$_2$. We adopted the Solar abundances 
from Morton (2003) based on meteoritic data from Grevesse \& Sauval (2002).

The characteristics of the sample are summarized in Table~\ref{summarytab}. The 
H~{\sc i} column densities and most of metallicities are from the compilation 
by Ledoux et al. (2006a). Slight differences with previously published
values, e.g. Ledoux et al. (2003), are due to the use of different damping
coefficients for H~{\sc i}. The only known $z>1.8$ H$_2$ bearing DLA system 
out of this sample is the system at $z_{\rm abs}$~=~2.337 toward Q~1232+082
(Srianand et al. 2000).


\begin{table*}
\caption{Metal and molecular content of high-metallicity DLA/sub-DLA systems at $z_{\rm abs}>1.8$}
\begin{tabular}{llllllccll}
\hline
\hline
QSO & $z_{\rm em}$ & $z_{\rm abs}$ & $\log N($H\,{\sc i}$)$ $^1$ & [X/H$]$ $^1$ & X  & \multicolumn{2}{c}{$\log N($H$_2)$ $^2$} & $\log f$ $^3$ & References\\
    &              &               &                             &              &    &            $J=0$ & $J=1$                 &               &           \\
\hline
Q\,0013$-$004  &  2.09  &  1.973  &  $20.83\pm 0.05$  &  $-0.59\pm 0.05$  &  Zn  &  \multicolumn{2}{c}{17.72--20.00}            &$ -1.68^{+1.07}_{-1.18}$ & $a$\\ 
Q\,0112$-$306  &  2.99  &  2.702  &  $20.30\pm 0.10$  &	 $-0.49\pm 0.11$  &  Si  &  $<14.1$ & $<14.0$                           &$<-5.55$ & $b$   \\
Q\,0216$+$080  &  2.99  &  2.293  &  $20.50\pm 0.10$  &	 $-0.70\pm 0.11$  &  Zn  &  $<14.3$ & $<14.5$                           &$<-5.39$ & $c$         \\
Q\,0347$-$383  &  3.22  &  3.025  &  $20.73\pm 0.05$  &	 $-1.17\pm 0.07$  &  Zn  &  \multicolumn{2}{c}{$14.53^{+0.06}_{-0.06}$}   &$ -5.90^{+0.11}_{-0.11}$ & $b$   \\
Q\,0405$-$443  &  3.02  &  2.595  &  $21.05\pm 0.10$  &	 $-1.12\pm 0.10$  &  Zn  &  \multicolumn{2}{c}{$18.14^{+0.07}_{-0.10}$}   &$ -2.61^{+0.17}_{-0.20}$ & $b,d$\\ 
Q\,0458$-$020  &  2.29  &  2.040  &  $21.70\pm 0.10$  &  $-1.22\pm 0.10$  &  Zn  &  $<14.6$ & $<14.6$                           &$<-6.40$ & $e$ \\
Q\,0528$-$250  &  2.77  &  2.811  &  $21.35\pm 0.07$  &  $-0.91\pm 0.07$  &  Zn  &  \multicolumn{2}{c}{$18.22^{+0.11}_{-0.12}$}   &$ -2.83^{+0.18}_{-0.19}$ & $b,d$ \\
Q\,0551$-$366  &  2.32  &  1.962  &  $20.70\pm 0.08$  &	 $-0.35\pm 0.08$  &  Zn  &  \multicolumn{2}{c}{$17.42^{+0.45}_{-0.73}$}   &$ -2.98^{+0.53}_{-0.81}$ & $f$ \\
Q\,1037$-$270  &  2.23  &  2.139  &  $19.70\pm 0.05$  &  $-0.31\pm 0.05$  &  Zn  &  $<14.0$ & $<14.1$                           &$<-5.05$ & $g$ \\
Q\,1209$+$093  &  3.30  &  2.584  &  $21.40\pm 0.10$  &	 $-1.01\pm 0.10$  &  Zn  &  $<14.9$ & $<15.1$                           &$<-5.69$ & $c$ \\  
Q\,1441$+$276  &  4.42  &  4.224  &  $20.95\pm 0.10$  &	 $-0.63\pm 0.10$  &   S  &  \multicolumn{2}{c}{$18.28^{+0.08}_{-0.07}$}   &$ -2.38^{+0.18}_{-0.17}$ & $h$ \\
Q\,1444$+$014  &  2.21  &  2.087  &  $20.25\pm 0.07$  &	 $-0.80\pm 0.09$  &  Zn  &  \multicolumn{2}{c}{$18.16^{+0.14}_{-0.12}$}   &$ -1.80^{+0.21}_{-0.19}$ & $b$ \\
Q\,2116$-$358  &  2.34  &  1.996  &  $20.10\pm 0.07$  &	 $-0.34\pm 0.11$  &  Zn  &  $<14.5$ & $<14.8$                           &$<-4.75$ & $c$\\  
Q\,2206$-$199  &  2.56  &  1.921  &  $20.67\pm 0.05$  &	 $-0.54\pm 0.05$  &  Zn  &  $<14.4$ & $<14.7$                           &$<-5.44$ & $c$ \\    
Q\,2230$+$025  &  2.15  &  1.864  &  $20.90\pm 0.10$  &	 $-0.81\pm 0.10$  &   S  &  $<15.4$ & $<15.4$                           &$<-4.80$ & $c$ \\
Q\,2243$-$605  &  3.01  &  2.331  &  $20.65\pm 0.05$  &	 $-0.85\pm 0.05$  &  Zn  &  $<13.8$ & $<13.9$                           &$<-6.15$ & $c$ \\ 
Q\,2343$+$125  &  2.51  &  2.431  &  $20.40\pm 0.07$  &	 $-0.89\pm 0.08$  &  Zn  &  \multicolumn{2}{c}{$13.69^{+0.09}_{-0.09}$}   &$ -6.41^{+0.16}_{-0.16}$ & $c$ \\
Q\,2348$-$011  &  3.01  &  2.426  &  $20.50\pm 0.10$  &	 $-0.60\pm 0.11$  &   S  &  \multicolumn{2}{c}{$18.45^{+0.27}_{-0.26}$}   &$ -1.76^{+0.37}_{-0.36}$ & $c$\\ 
\hline
\end{tabular}
\label{summarytab}
\flushleft
$^1$ Neutral hydrogen column densities and metallicities relative to
Solar, [X/H$]\equiv\log [N($X$)/N($H$)]-\log [N($X$)/N($H$)]_\odot$
(with ${\rm X}={\rm Zn}$ as the reference element when Zn\,{\sc ii}
is detected, or else either S or Si) are from Ledoux et al. (2006a).\\
$^2$ Molecular hydrogen column densities are summed up over all $J$
levels in case of detection and upper limits for ${\rm J}=0$ and
${\rm J}=1$ are given in case of non-detection.\\
$^3$ Molecular fraction, $f=2N($H$_2)/(2N($H$_2)+N($H\,{\sc i}$))$.\\
$a$ Petitjean et al. (2002), $b$ Ledoux et al. (2003), $c$ This work, $d$ Srianand et al. (2005),
$e$ Heinm\"uller et al. (2006), $f$ Ledoux et al. (2002b), $g$ Srianand \& Petitjean (2001),
$h$ Ledoux et al. (2006b)
\end{table*}


\begin{figure}[!ht]
  \begin{center}
\includegraphics[width=10.0cm,bb=68 710 399 766,clip=]{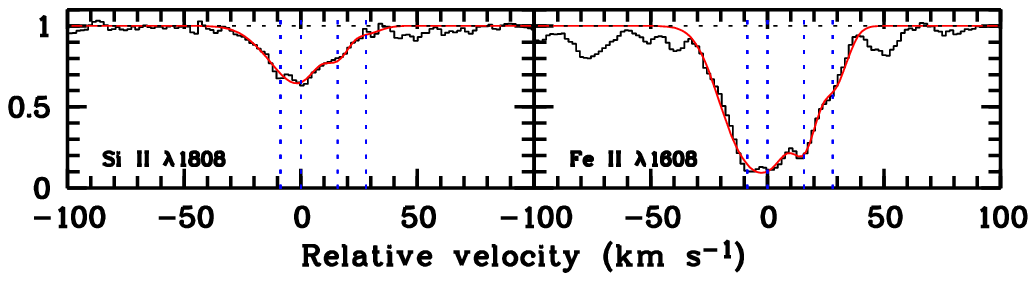}\\
\includegraphics[width=10.0cm,bb=68 483 399 766,clip=]{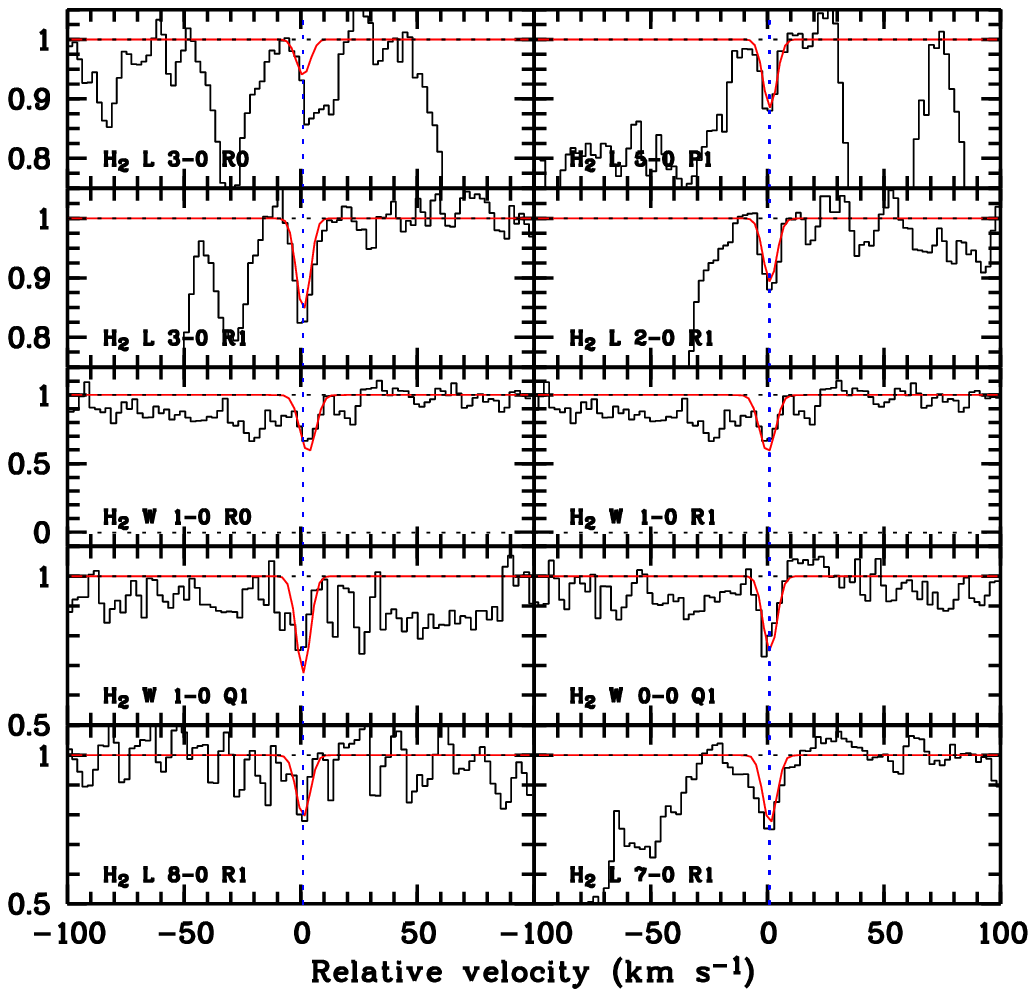}\\
   \caption{Velocity plots of observed H$_2$ absorption lines from
the J~=~0 and J~=~1 rotational levels 
at $z_{\rm abs}$~=~2.43127 toward Q\,2343$+$125. Profiles of associated Si~{\sc ii} and
Fe~{\sc ii} absorptions are shown in the upper panels. The best-fitting model
to the system is overplotted; the location of Voigt-profile sub-components are 
shown as vertical dashed lines. }
\label{q2343h2}
  \end{center}
\end{figure}

\section{Two new detections}

We report two new detections of H$_2$ in DLAs from our new observations.
Detail analysis and interpretation of the physical conditions in these two
DLAs are out of the scope of the present work and will be described in a subsequent 
paper.

\subsection{Q\,2343$+$125, $z_{\rm abs} = 2.431$}





This DLA system has first been studied by Sargent et al. (1988). High resolution data
have been described by Lu et al. (1996), D'Odorico et al. (2002) and Dessauges-Zavadsky
et al. (2004). The profile of the metal lines is spread over more than 250~km~s$^{-1}$
from $z_{\rm abs}$~=~2.4283 to 2.4313 but the strongest component is centered 
at $z_{\rm abs} = 2.43127$ corresponding to the red edge of the above redshift range.
From Voigt--profile fitting to the H~{\sc i} Lyman--$\alpha$, $\beta$ and $\gamma$ lines, 
we find that the damped Lyman-$\alpha$ line is centered at  $z_{\rm abs} = 2.431$ and the
column density is log $N$(H~{\sc i}) = $20.40 \pm 0.07$, consistent with previous measurement by 
D'Odorico et al. (2002; log $N$(H~{\sc i}) = $20.35 \pm 0.05$). 
We use Zn as the reference species for metallicity measurement and find
[Zn/H]~=~$-$0.89$\pm$0.08. This is consistent with previous findings.
Absorption from the J~=~1 and probably from the J~=~0 rotational levels 
of H$_2$ is detected in this system at $z_{\rm abs} = 2.43127$ (see Fig.~\ref{q2343h2}).
The optically thin H$_2$ absorption lines are very weak, i.e. close to but above the 
3$\sigma$ detection limit.
A very careful normalization of the spectrum has been performed, 
adjusting the continuum while fitting the lines.
%
The best-fitting consistent model for H$_2$ is shown in Fig.~\ref{q2343h2}.
The total H$_2$ column density, integrated over the J~=~0 and 1 levels, is 
estimated to be log $N$(H$_2$) = $13.69 \pm 0.09$ (12.97$\pm$0.04 and 13.60$\pm$0.10
for J~=~0 and 1 respectively). This leads to the smallest molecular fraction observed
up to now, 
log~$f$~=~$-6.41_{-0.16}^{+0.16}$.
We also derive an upper limit on the detection of absorption from the J~=~2 level,
log $N$(H$_2$-J=2) $<$ $13.1$ at the 3$\sigma$ level.

%

\subsection{Q\,2348$-$011, $z_{\rm abs} = 2.426$}

\begin{figure}[!ht]
  \begin{center}
\includegraphics[width=10.0cm,bb=68 659 392 766,clip=]{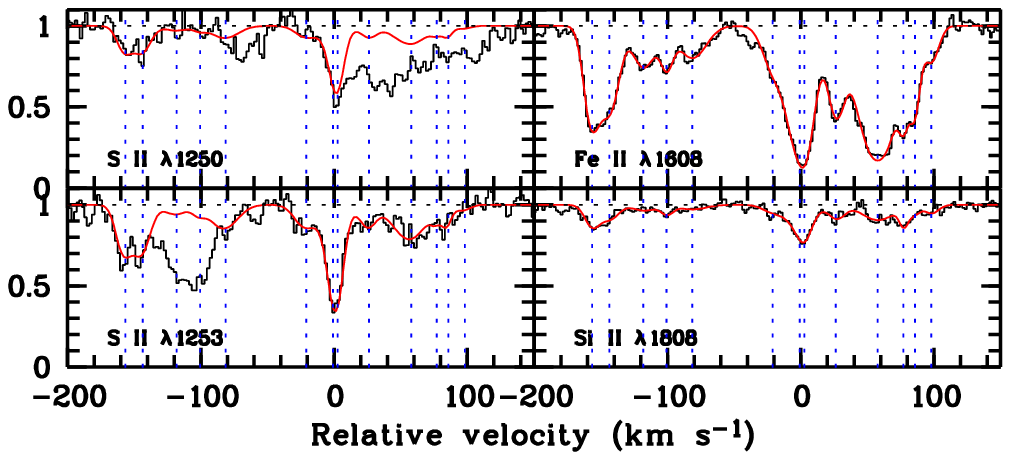}\\
\includegraphics[width=10.0cm,bb=68 688 392 767,clip=]{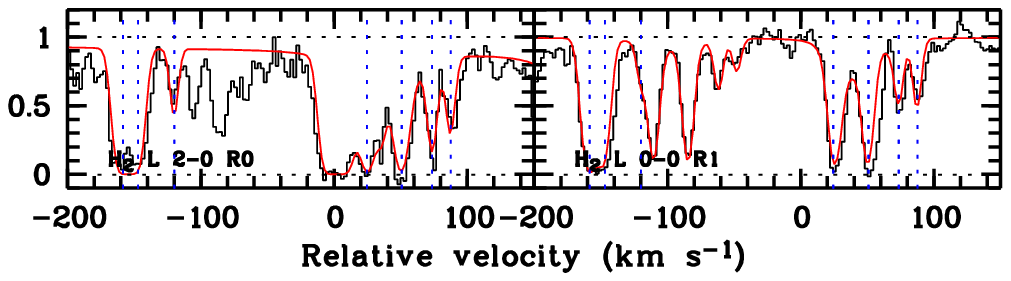}\\
\caption{Absorption profiles at $z_{\rm abs}$~=~2.4263 (taken as
zero of the velocity scale) toward Q\,2348$-$011. The positions of the seven
(respectively 13) components needed to fit the H$_2$ (respectively metal line)
profiles are indicated by vertical dashed lines. The resulting best-fitting model
to the system is overplotted. }
 \label{q2348}
 \end{center}
\end{figure}

There are two DLA systems at $z_{\rm abs}=2.426$ and $z_{\rm abs} = 2.615$ toward 
Q\,2348$-$011, with total neutral hydrogen column densities of respectively 
log $N$(H~{\sc i}) = $20.50 \pm 0.10$ and log $N$(H~{\sc i}) = $21.30 \pm 0.08$. Conspicuous 
H$_2$ absorptions are detected in the  $z_{\rm abs}\simeq 2.426$ DLA system, the only system 
to be considered here as displaying a high metallicity (see Fig.~\ref{q2348}).
The molecular lines are very numerous and strong but the spectral resolution of our data is 
high enough to allow unambiguous detection and accurate determination of the line parameters.
%
Seven H$_2$ components spread over about 300~km~s$^{-1}$ were used for the H$_2$ fit. 
It is interesting to note that the strongest metal component (at $V$~=~0~km~s$^{-1}$
in Fig.~\ref{q2348}) has no 
associated H$_2$ absorption. Strong H$_2$ absorption is seen at $V$~$\sim$~$-$150~ and +50~km~s$^{-1}$.
All seven molecular components have associated C~{\sc i} absorption. However, additional components
are needed to fit the metal absorption lines: 9 components for C~{\sc i} and 13 components
for the singly ionized species. 
H$_2$ absorption from the rotational levels J~=~0 to 5 are unambiguously detected. 
The total H$_2$ column density integrated 
over all rotational levels is log~$N$(H$_2$)~=~18.45$^{+0.27}_{-0.26}$, 
corresponding to a molecular fraction log~$f$~=~$-1.76_{-0.36}^{+0.37}$. We also derive an upper limit 
on the column density of HD molecules, leading to log~$N$(HD)/$N$(H$_2$) $<$ -3.3. 


\section{Metallicity as a criterion for the presence of molecular hydrogen}

\begin{figure}[!ht]
  \includegraphics[width=8.2cm,height=9.5cm,bb=0 0 574 726, clip,angle=90]{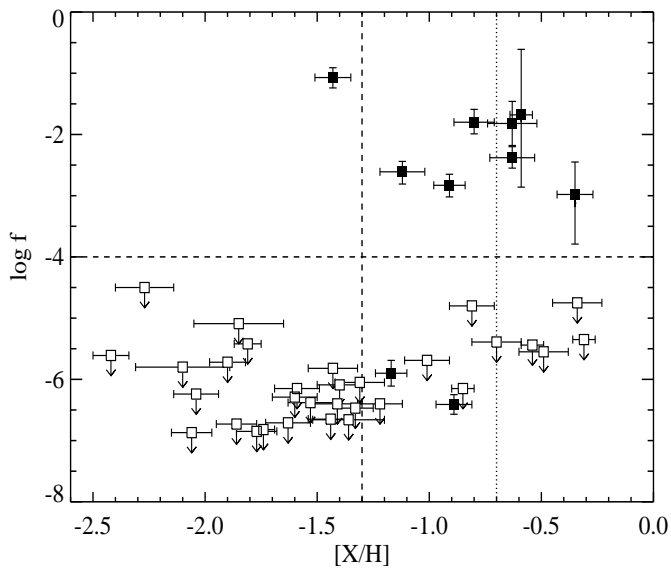}
   \caption{Logarithm of the molecular fraction, $f$~=~2$N$(H$_2$)/(2$N$(H$_2$)~+~$N$(H~{\sc i})),
versus metallicities, [X/H] = log~$N$(X)/$N$(H)$-$log(X/H)$_{\odot}$ and X either Zn, S or Si, 
in DLAs from the sample described in this paper ([X/H]~$>$~$-1.3$, see Table~1) and
the sample of Ledoux et al. (2003, [X/H]~$<$~$-1.3$). Filled squares indicate
systems in which H$_2$ is detected. Dashed lines indicate the limits used in the text
(log~$f$~=~$-$4; [X/H]~=~$-$1.3). The dotted line indicate the median of the
high-metallicity sample ([X/H]~=~$-$0.7).}
\label{sample}
\end{figure}

From Table 1, it can be seen that H$_2$ is detected in nine high-metallicity 
systems out of 18. However, for two of the detections, the corresponding 
value of log~$f$ is smaller than most of the upper limits derived for
other systems. All upper limits are smaller than $-$4.5 and all detections
larger than these upper limits are larger than $-$3. We therefore define a system
with large (respectively small) H$_2$ content if log~$f$ is larger (respectively smaller) than 
$-$4.

In Fig.~\ref{sample} we plot the molecular fraction, log~$f$, versus metallicity, [X/H],
for our representative sample of DLAs with [X/H]~$>$~$-1.3$ (18 measurements summarized 
in Table~1) and measurements by Ledoux et al. (2003) for [X/H]~$<$~$-1.3$ (23 measurements).
The log~$f$ distribution is bimodal with an apparent gap
in the range $-5$~$<$~log~$f$~$<$~$-3.5$ justifying the above classification
of systems.
Note that this jump in log~$f$ has already been noticed before by Ledoux et al. (2003) and 
is similar to what is seen in our Galaxy (Savage et al. 1977; see also Srianand et al. 2005).
%
It is apparent that the fraction of systems with 
molecular fraction log~$f$~$>$~$-4$ increases with increasing metallicity. It 
is only $\sim$5\% for [X/H]~$<$~$-1.3$
when it is $\sim$39\% for [X/H]~$>$~$-1.3$. This fraction is even larger, 
50\%, for [X/H]~$>$~$-0.7$ which is the median metallicity for systems
with [X/H]~$>$~$-1.3$. In addition, all systems with [X/H]~$<$~$-1.5$
have  log~$f$~$<$~$-4.5$.

We conclude that metallicity is an important criterion for the presence of molecular 
hydrogen in DLAs. This may not be surprizing as the correlation 
between metallicity and depletion of metals onto dust grains (Ledoux et al. 2003)
implies that larger metallicity means larger dust content and therefore
larger H$_2$ formation rate. In addition, the presence of dust implies a
larger 
absorption of UV photons that usually dissociate the molecule. More generally, 
Ledoux et al. (2006a) have
shown that a correlation exists between metallicity and velocity width in DLAs.
If the latter kinematic parameter is interpreted as reflecting the mass
of the DM halo associated with the absorbing object, then DLAs with higher
metallicity are associated with objects of larger mass in which star formation
could be enhanced.
All this makes it arguable that, in DLAs, star-formation activity is probably correlated with 
the molecular fraction (Hirashita \& Ferrara 2005). 
It is therefore of first importance to survey a large number of DLA systems 
to define better their molecular content and use this information 
to derive the physical properties of the gas and the amount of star-formation
occuring in the associated objects.


\begin{acknowledgements}
PPJ thanks ESO for an invitation to stay at the ESO headquarters in Chile where part of
this work was completed. RS
and PPJ gratefully acknowledge support from the Indo-French Centre for
the Promotion of Advanced Research (Centre Franco-Indien pour la Promotion
de la Recherche Avanc\'ee) under contract 3004-3. PN is supported by an
ESO student fellowship.
\end{acknowledgements}

\end{document}